\documentclass[12pt]{iopart}

\usepackage{graphicx}   % for figures
\begin{document}
\title[Energy-momentum density and pressure in a relativistic ideal gas] {Energy-momentum density and pressure relations for a relativistic ideal gas with a bulk motion} 
\author{Ashok K. Singal}
\address{Astronomy and Astrophysics Division, Physical Research Laboratory,
Navrangpura, Ahmedabad - 380 009, India }
\ead{ashokkumar.singal@gmail.com}
\vspace{10pt}
\begin{indented}
\item[]\today
\end{indented}
%--------------------------------------------------------------------
\begin{abstract}
We derive here, from first principles, the energy-momentum densities of a perfect fluid, in the form of an ideal molecular gas, in an inertial frame where the fluid possesses a bulk motion. We begin  from the simple expressions for the energy density and pressure of a perfect fluid in the rest frame of the fluid, where the fluid constituents (gas molecules) may possess a random motion, but no bulk motion. From a Lorentz transformation of the velocity vectors of  molecules, moving along different directions in the rest frame of the fluid, we compute their energy-momentum vectors and number densities in an inertial frame moving with respect to the rest frame of the liquid. From that we arrive at the energy-momentum density of the fluid in a frame where it has a bulk motion. This way we explicitly demonstrate how a couple of curious pressure-dependent terms make appearance in the energy-momentum density of a perfect fluid having a bulk motion. In addition to an ideal molecular gas, we compute the energy-momentum density for a photon gas also, which of course matches with the energy-momentum density expression obtained for a molecular gas having ultra-relativistic random motion.
\end{abstract}
%PhySH: Astrophysical fluid dynamics; Electromagnetism; Special Relativity
%--------------------------
\section{Introduction}
One of the simplest examples of a perfect fluid is an ideal gas, where the constituent molecules of the gas are assumed to be structureless finite masses of negligible dimensions that may have head-on elastic collisions with each other as well as with the walls of the container but have otherwise no interaction. Thus there are no side-ways forces, thereby no drag or viscosity as well as no sheer stresses and there is no conduction of heat. Such a fluid is described completely by the energy density and pressure in an inertial frame, called the rest frame, where the pressure is isotropic, and the stress-energy tensor has a simple, diagonal form. Perfect fluids are used in making relativistic models of the interior of a star where an idealized distributions of matter may be assumed \cite{MTW73,del19}. Perfect fluids are also employed in studying the evolution of isotropic universe models using the general theory of relativity \cite{MTW73,BW17}.

In an inertial reference frame, where the fluid may have a bulk motion, the stress-energy tensor may possess off-diagonal terms too. Now in the component expressions of the stress-energy tensor of a perfect fluid with a bulk motion, there are a couple of pressure-dependent terms, that make explicit contributions to the energy and momentum densities, whose genesis remains mired in mystery. It is not very apparent how the pressure within the medium, in the presence of a bulk motion of the fluid, gives rise to these terms. Here we shall attempt to understand  how the bulk motion of the fluid, causes the random motion of the fluid constituents in the rest frame, to give rise to a couple of pressure-dependent terms in the energy and momentum densities of the fluid elements. 

These questions have recently been addressed elsewhere too \cite{84}, where it was demonstrated that these pressure-dependent terms in the energy and momentum densities arise from two subtle factors: (i) When the liquid is accelerated from the rest frame, say ${\cal S}_0$, to another frame, say ${\cal S}$, where a fluid element thus attains a bulk motion, work gets done against pressure during an increasing Lorentz contraction of the fluid element. (ii) During the movement of a fluid element, work is continuously being done {\em on} the fluid element at its trailing end against pressure, while an equal and opposite work is being done {\em by} the fluid element at its  leading end. Thus the energy gained by the fluid element at its trailing end is being delivered by the fluid element at its leading end. Even though there is no net gain or loss of energy, yet due to this, there is a continuous flow of energy across the fluid element between its two ends, which results in an additional contribution to the momentum of the fluid element. This latter term was in fact shown to have given rise to the intriguing, century old, 4/3 problem \cite{29,1,2,15} in the electromagnetic momentum of a moving charge. Further, in the apparent paradox of there being nil  electromagnetic momentum in a moving electrically charged parallel-plate capacitor, carrying a finite electromagnetic field energy during its motion along the plate separation, added to the fact that the field energy paradoxically decreases with an increase in the system velocity, it was the contribution of these very terms to the energy-momentum of the electromagnetic system that was shown to be responsible for a successful resolution of these paradoxes \cite{84,15,31}. 

Even though the work done during a Lorentz contraction as well as the momentum contribution of differential work being done on the ends of a moving fluid element explained in an unambiguous manner how during the acceleration of the fluid to achieve a bulk motion, the energy and momentum densities get contributions from the pressure within the fluid, it would nevertheless be desirable if one was able to derive these pressure-dependent terms in the energy and momentum densities, from first principles, using Lorentz transformations, directly in the frame ${\cal S}$ itself, where the fluid has achieved a bulk motion. It is true that using the tensor notation leads to results in a rather quick and an easy manner, still a somewhat laborious exercise using velocity transformations from one Lorentz frame to another and the thereby computation of energy-momentum densities could be quite instructive.

It seems that for this purpose one might require some specific models of the fluid. We endeavour to do so here for two commonly used models of a relativistic fluid, (i) an ideal molecular gas (ii) a photon gas.
While the molecular gas could have either a non-relativistic or a relativistic random motion, the photon gas is always relativistic. The bulk motion, which is independent of the random motion of the gas, as the bulk motion could depend upon our choice of the reference frame, and could also be either non-relativistic or relativistic. This way we can explore various scenarios of relative magnitudes of the bulk motion and the random motion. 
%-----------------------------------------------
\section{Energy-momentum density of a perfect fluid with a bulk motion} 
Let ${\cal S}_0$ be the rest frame of the perfect fluid, then the energy-momentum tensor in  ${\cal S}_0$ is of a diagonal form \cite {MTW73,84,SC85,LL98,DS98} 
\begin{eqnarray}
\label{eq:85.1}
T'^{00} = \rho_0, \:\:T'^{11} =T'^{22}=T'^{33} =  p\,,
\end{eqnarray}
with all non-diagonal terms zero, implying an energy density $\rho_0$ and an isotropic pressure $p$, but a nil momentum density (by definition). It is important to note that $\rho_0$ here includes not only the rest mass energy of the fluid molecules, but also their kinetic energy, which for instance, in the ideal gas model of the fluid, arises due to the random motion of the gas molecules that also generates pressure in ${\cal S}_0$.

In the lab frame ${\cal S}$, where the perfect fluid has a bulk motion $v_0$, say along the x-axis, with a Lorentz factor $\gamma_0=(1-v_0^2/c^2)^{-1/2}$, the energy and momentum densities are given by the components $T^{00}$ and $T^{10}=T^{01}$ of the  energy-momentum tensor, \cite {MTW73,84,SC85,LL98,DS98} as
\begin{eqnarray}
\label{eq:85.4c}
\rho=T^{00} = \gamma_0^2 \left(\rho_0 +p {v_0^2\over c^2}\right)\:,
\end{eqnarray}
\begin{eqnarray}
\label{eq:85.5c}
\sigma=T^{10}/c =\gamma_0^2\left(\rho_0 +p\right) {v_0\over c^2}\: ,
\end{eqnarray}

Here a perplexing question could be raised. Why does the energy density 
$\rho$ in frame ${\cal S}$ depend upon pressure $p$ (Eq.~(\ref{eq:85.4c})), when the random motion of molecules in frame ${\cal S}_0$ is already accounted for by the kinetic energy, which is included in $\rho_0$. How come the random motion enters twice in this expression, once through the kinetic energy of the random motion and the second time through the pressure term which again results from the random motion? In fact, one would intuitively expect $\rho = \gamma_0^2\rho_0$ (the extra $\gamma_0$ factor simply due to the Lorentz contraction of the liquid volume, as seen in frame ${\cal S}$).

Equally intriguing is the presence of $p$ in $\sigma$, the momentum density (Eq.~(\ref{eq:85.5c})). With $\rho_0 {/ c^2}$ being the mass density of the system in rest frame ${\cal S}_0$ of the fluid, one would have expected the momentum density to be simply $\sigma= \gamma_0^2\rho_0 v{/ c^2}$. Why does an additional term ($\gamma_0^2 p {v_0/ c^2}$), proportional to pressure, appear in the expression for $\sigma$?

%---------------------------------------------------------------------
%---------------------------------------------------------
\section{An ideal molecular gas as a perfect fluid}
Let us assume an ideal gas, enclosed inside a container of volume $V_0$, having a uniform number density $n_0$ of molecules, each of rest mass $m_0$ and moving with a velocity $v$ in a random direction, as seen in the rest frame ${\cal S}_0$.  Let $\gamma= ({1-{ {v^{2}}/{c^{2}}}})^{-1/2}$ be the Lorentz factor of the random motion of molecules.
The pressure for the gas is related to the energy density as (Appendix A, Eq.~(\ref{eq:85.a4}))
\begin{eqnarray}
\label{eq:85.41}
    {\displaystyle p={\frac {n_0m_0\gamma{ {v^{2}}}}{3}}
    ={\frac {\rho_0{ {v^{2}}}}{3c^2}}\,,}
\end{eqnarray}
where $\rho_0=n_0m_0 c^2\gamma$ is the relativistic energy density of the gas that includes the rest mass energy of molecules as well as the kinetic energy due to their random motion in frame ${\cal S}_0$. If the motion of molecules in the gas is relativistic with $v\sim c$, the pressure $p$ would be comparable to the energy density $\rho_0$. 

As seen in the lab frame ${\cal S}$, the container of fluid moves with a velocity $v_0$ along  the x-axis, with $\gamma_0= ({1-{ {v_0^{2}}/{c^{2}}}})^{-1/2}$ as the corresponding Lorentz factor.
Now a molecule moving along $\theta_0$ with respect to the x$_0$-axis, as seen in  ${\cal S}_0$, will be moving with a velocity component, $V_{\rm x}$, parallel 
%and perpendicular 
to the x-axis in the lab frame ${\cal S}$, given by \cite{DS98,KKR73,RI06}
\begin{eqnarray}
\label{eq:85.42}
{\displaystyle V_{\rm x}={\frac {v_{\rm x}+v_0}{1+{ {v_{\rm x}v_0}/{c^{2}}}}}\,,}
 \end{eqnarray}
Writing $V_{\rm x}=V \cos\theta$ and $v_{\rm x}=v \cos\theta_0$, we can write Eq.~(\ref{eq:85.42}), as
\begin{eqnarray}
\label{eq:85.47}
{\displaystyle V\cos\theta={\frac {v \cos\theta_0+v_0}{1+{ {v_0v \cos\theta_0}/{c^{2}}}}}\,,}
 \end{eqnarray}
from which we get
\begin{eqnarray}
\label{eq:85.47.1}
{\displaystyle V\cos\theta-v_0={\frac {v \cos\theta_0}{\gamma^2_0(1+{ {v_0v \cos\theta_0}/{c^{2}}})}}\,.}
 \end{eqnarray}
%---------------------------------------------
\begin{figure}[t]
\begin{center}
%\scalebox{0.9}{\includegraphics{singal_ashokkfig01.jpg}}
\scalebox{0.9}{\includegraphics{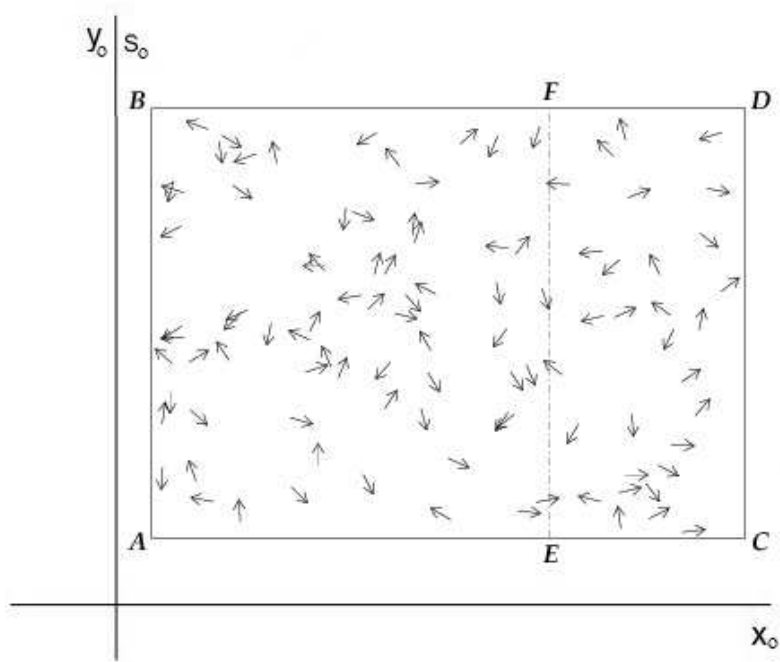}}
%\scalebox{0.9}{\includegraphics{fig1.eps}}
\end{center}
\caption{A box bound by walls, $AB$, $CD$, $AC$ and $BD$ (the depth along the z-axis suppressed in the diagram), containing ideal gas, is stationary in frame ${\cal S}_0$. The gas constituents may have a random motion with an isotropic distribution, and thus no bulk motion, in frame ${\cal S}_0$. $EF$ (dashed line) is a cross section of the container, at some arbitrary position along $x_0$ direction.}
\end{figure}
%---------------------------------------------

Now even if the motion of molecules in the frame ${\cal S}_0$ has an isotropic distribution, it would not be so in the frame ${\cal S}$ (Fig.~2). This curious fact arises because all those  molecules that may have velocity component towards right (that is along positive $x$ direction) in frame  ${\cal S}_0$, will also move towards right in frame ${\cal S}$ too and with somewhat larger magnitudes. In addition, all those molecules will also be moving towards right in the frame ${\cal S}$ which were moving in frame ${\cal S}_0$ towards left but with velocity magnitudes smaller that $v_0$, the velocity of the frame ${\cal S}_0$ with respect to  the frame ${\cal S}$. One has to also consider that the container walls, $AB$ and $CD$, though at rest in frame ${\cal S}_0$, are in motion in  ${\cal S}$. Therefore, successive molecules bouncing off the wall $CD$, and moving towards left after that, have larger separations, due to the meanwhile movement of the wall towards right, as compared to the separation between molecules that are bounced off towards right by the wall $AB$. As a result there may be, on the average, a larger number of molecules moving towards right than those moving towards left, as seen in frame  ${\cal S}$. The relative densities for molecules moving at different angles $\theta$ with respect to the x-axis, can be determined in the following way.
%---------------------------------------------
\begin{figure}[t]
\begin{center}
%includegraphics[width=\columnwidth]{singal_ashokkfig02.jpg}
\scalebox{0.9}{\includegraphics{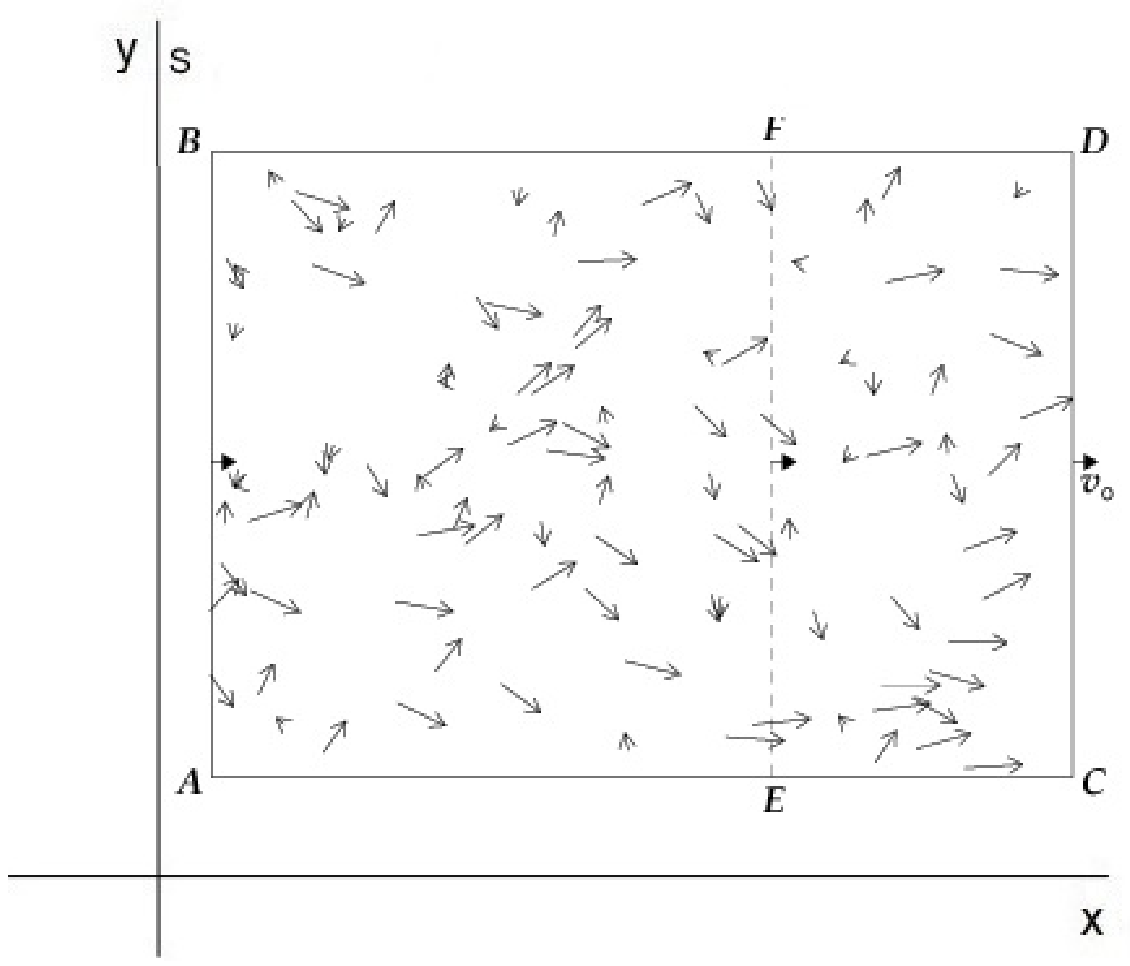}}
%\scalebox{0.9}{\includegraphics{fig2.eps}}
\end{center}
\caption{The box $ABDC$, at test in frame ${\cal S}_0$, is moving with velocity ${v}_0$ along the x-axis with respect to the lab frame ${\cal S}$. Even the cross section, $EF$ (dashed line), of the container is moving with velocity ${v}_0$ in frame ${\cal S}$. Figure shows  in frame ${\cal S}$, the Lorentz transformed velocities of the gas molecules, which had an isotropic distribution in frame ${\cal S}_0$.}
\end{figure}
%---------------------------------------------

The time $\tau$ that a molecule moving with velocity $V_{\rm x}$ in frame ${\cal S}$ would take between successive collision with two opposite container walls $AB$ and $CD$, that are normal to the x-axis and moving towards right with speed $v_0$ (Fig.~2), is given by  
\begin{eqnarray}
\label{eq:85.45}
V_{\rm x}\tau =\frac{l}{\gamma_0}+v_0\tau\,,
\end{eqnarray}
where length $l$ of the arm AC of the container in frame ${\cal S}_0$ is Lorentz contracted in frame ${\cal S}$, along the $x$ direction, by a factor $\gamma_0= ({1-{ {v_0^{2}}/{c^{2}}}})^{-1/2}$, and the wall $CD$ moves a distance $v_0\tau$ towards right in time $\tau$.
Thus the time interval between successive collisions with the container walls
\begin{eqnarray}
\label{eq:85.46}
\tau =\frac{l}{\gamma_0(V_{\rm x}-v_0)}\,,
\end{eqnarray}
in frame ${\cal S}$ will be longer than $\tau_0=l/v_{\rm x}$ in frame ${\cal S}_0$ and the corresponding molecule will be spending relatively more time after bouncing off the container wall $AB$ by a factor, $v_{\rm x}/{\gamma_0}(V_{\rm x}-v_0)$, before a change in $V_{\rm x}$ takes place because of a collision with the container wall $CD$. Therefore the number density of molecules moving with $V_{\rm x}$, as seen in frame ${\cal S}$, will be relatively higher than those seen  moving with $v_{\rm x}$ in frame ${\cal S}_0$, by a factor 
\begin{eqnarray}
\label{eq:85.46a}
\frac{\tau}{\tau_0}=\frac{v_{\rm x}}{\gamma_0(V_{\rm x}-v_0)}\,.
\end{eqnarray}
Therefore the number density in frame ${\cal S}$ will not be isotropic, even though in frame ${\cal S}_0$ the number densities of molecules with velocity components in all directions were uniformly distributed.

With the help of Eq.~(\ref{eq:85.47.1}), we get
\begin{eqnarray}
\label{eq:85.48}
\frac{\tau}{\tau_0}= \frac{v\cos\theta_0}{\gamma_0(V\cos\theta-v_0)}=\gamma_0\left(1+\frac{v_0v\cos\theta_0}{c^2}\right)={\cal D}\,(\rm say) \:,
\end{eqnarray}
${\cal D}$ is akin to the Doppler factor $\delta$ for photons (see the next Section). 
Thus, the number densities of molecules within a given liquid element in frame ${\cal S}$, will be ${\cal D}$ times that in frame ${\cal S}_0$, which is different from the mere Lorentz contraction factor, $\gamma_0$, of the volume of liquid element.

Here we should point out that for molecules moving towards $-x$ direction in the frame ${\cal S}_0$, i.e., for molecules with $\theta_0 > \pi/2$, $V_{\rm x}<v_0$ and both 
${\tau}$ and ${\tau_0}$ are negative. What really happens here is that instead of the time interval between ($n$)th and ($n+1$)th collision, one gets in such cases the time interval between ($n$)th and ($n-1$)th collision. However the ratio ${\tau}/{\tau_0}$ still correctly gives  ${\cal D}$, the ratio of number densities in frames ${\cal S}$ and ${\cal S}_0$ for the corresponding molecules.

Let $\Gamma= ({1-{ {V^{2}}/{c^{2}}}})^{-1/2}$ be the Lorentz factor of the motion of the molecule in the lab frame ${\cal S}$. Then from a Lorentz transformation of the velocity 4-vector $\gamma (c,v_{\rm x},v_{\rm y},v_{\rm z})$ in ${\cal S}_0$ to $\Gamma (c,V_{\rm x},V_{\rm y},V_{\rm z})$ in  ${\cal S}$, we have
\begin{eqnarray}
\label{eq:85.43}
\Gamma=\gamma_0\gamma\left(1+\frac{v_0v_{\rm x}}{c^2}\right)=\gamma{\cal D}\,,
\end{eqnarray}
implying, the energy of a molecule in frame ${\cal S}$ will be ${\cal D}$ times that in frame ${\cal S}_0$.

Here a question could be still baffling -- While it could be readily understood that the energy of a molecule moving along $\theta_0$  in frame ${\cal S}_0$, could be different in frame ${\cal S}$ from that in frame ${\cal S}_0$ (Eq.~(\ref{eq:85.43})), how come the number densities of molecules in different directions within a given liquid element vary in frame ${\cal S}$, beyond the Lorentz contraction factor, $\gamma_0$, of the volume of liquid element? After all  they remain uniform in frame ${\cal S}_0$, not only in all directions but also at all times. For instance, we could select a cross section, say $EF$ normal to the x-axis (Fig.~1) within the fluid, then in frame ${\cal S}_0$, with an equal number of molecules crossing EF from left to right as from right to left, we will see a zero net number-flux through $EF$, i.e. 
\begin{eqnarray}
\label{eq:85.49b}
\frac{n_0 v}{4\pi}\int_{0}^{\pi}  \cos\theta_0\, 2\pi \sin \theta_0 \: {\rm d}\theta_0=\frac{n_0 v}{2}\int_{0}^{\pi}  \cos\theta_0\, \sin \theta_0 \: {\rm d}\theta_0=0\,,
\end{eqnarray}
The same set of molecules, with a nil net number flux, as seen in  ${\cal S}_0$, should be seen in frame ${\cal S}$ too, crossing $EF$ in equal numbers in opposite directions simultaneously. 
How could it be possible if the number density of molecules moving towards right is different from that of the ones moving towards left?

Actually the  cross section $EF$, as seen in ${\cal S}$,  will be moving towards right with a speed $v_0$ (Fig.~2), and the molecules moving towards right, with number densities higher by factor ${\cal D}$, will have to catch up with the {\em receding away} cross section $EF$, while for molecules moving towards left, $EF$ also will be moving {\em towards} them, and in spite of their lesser densities, the number of molecules encountering $EF$ head on and crossing it from right to left need not be at a proportionally lower rate. Also, even if a molecule lying on left of $EF$ is moving towards right, but with a velocity $V_{\rm x} < v_0$, it would not be able to catch up with $EF$ to cross it from left to right. However, if a similar molecule, with a velocity $V_{\rm x} < v_0$,  lies on the right side of $EF$, then it could be overtaken by $EF$ moving towards right with velocity $v_0$, and such a molecule will be seen to pass through cross-section $EF$ from right to left, even though its velocity in ${\cal S}$ is towards right. As a result of these effects, the total number of molecules crossing $EF$ in opposite directions could be still equal, even in ${\cal S}$.
 
To verify this explicitly, we calculate the net number of molecules, ${\cal N}$, crossing per unit of time, through a moving cross section $EF$ that is moving towards right with speed $v_0$, as seen in ${\cal S}$. 
\begin{eqnarray}
\nonumber
{\cal N} &=&\frac{n_0}{4\pi}\int_{0}^{\pi} {\cal D} (V\cos\theta-v_0)\, 2\pi \sin \theta_0 \: {\rm d}\theta_0\,,
\end{eqnarray}
\begin{eqnarray}
\label{eq:85.49}
&=&\frac{n_0 v}{2\gamma_0}\int_{0}^{\pi}  \cos\theta_0 \sin \theta_0 \: {\rm d}\theta_0=0\,,
\end{eqnarray}
where we have used Eq.~(\ref{eq:85.48}). The appearance of $\gamma_0$ in the denominator on the right hand side in Eq.~(\ref{eq:85.49}), as compared to that in Eq.~(\ref{eq:85.49b}), merely indicates the time dilation (${\rm d}t=\gamma_0\,{\rm d}t_0$) for the number of molecules through the {\em moving} cross-section $EF$, per unit time, observed in frame $\cal S$. In any case, the net number crossing $EF$ is nil, implying the number of molecules crossing  $EF$ in opposite directions being equal, in frame $\cal S$ too.

Now, for computing the energy density of the gas in frame $\cal S$, we can make use of the fact that the isotropic distribution of the rest frame ${\cal S}_0$ has got modified, because for the molecules moving within an element of solid angle ${\rm d}\Omega_0=\sin \theta_0 \: {\rm d}\theta_0\:{\rm d}\phi$ around $\theta_0$ in frame ${\cal S}_0$, not only the number density gets modified by a factor ${\cal D}$, even the energy of individual molecules ($m_0 c^2\gamma$) in ${\cal S}_0$, is higher by a factor ${\cal D}$ in ${\cal S}$. Then an integration over the solid angle yields  for the energy density, $\rho$, as
\begin{eqnarray}
\nonumber
\rho &=&\frac{n_0m_0 c^2\gamma}{4\pi}\int_{0}^{2\pi}\int_{0}^{\pi} {\cal D}^2\, \sin \theta_0 \: {\rm d}\theta_0\:{\rm d}\phi=\frac{n_0m_0 c^2\gamma}{4\pi}\int_{0}^{\pi} {\cal D}^2\, 2\pi\sin \theta_0 \: {\rm d}\theta_0\,,\\
\nonumber
&=&\frac{\rho_0}{2}\int_{0}^{\pi} \gamma_0^2 \left(1+\frac{v_0v \cos \theta_0}{c^2}\right)^2\, \sin \theta_0 \: {\rm d}\theta_0\\
\label{eq:85.49a}
&=&\rho_0\gamma_0^2\left(1+\frac{v_0^2v^2}{3c^2}\right)=\gamma_0^2\left(\rho_0+pv_0^2\right)\,,
\end{eqnarray}
where we have used Eq.~(\ref{eq:85.41}) for $p$, the pressure. Here $\rho_0=n_0m_0 c^2\gamma$ is the relativistic energy density of the gas, that includes the rest mass energy density as well as the kinetic energy density of random motion of molecules in rest frame ${\cal S}_0$. 

The momentum of a molecule moving along $\theta_0$ in frame ${\cal S}_0$ is $m_0\Gamma\: V\cos\theta=m_0\gamma {\cal D}V\cos\theta$ in frame ${\cal S}$. Then proceeding in the same way as for the energy density, the momentum density is computed as
\begin{eqnarray}
\nonumber
\sigma &=&\frac{n_0m_0\gamma}{4\pi}\int_{0}^{\pi}  {\cal D}^2\, V\cos\theta\, 2\pi\sin \theta_0 \: {\rm d}\theta_0\,,\\
\nonumber
&=&\frac{\rho_0\gamma_0^2}{2 c^2}\int_{0}^{\pi}  \left(1+\frac{v_0v\cos \theta_0}{c^2}\right)^2V\cos\theta \sin \theta_0 \: {\rm d}\theta_0\\
\nonumber
&=&\frac{\rho_0\gamma_0^2}{2 c^2}\int_{0}^{\pi}  \left(1+\frac{v_0v\cos \theta_0}{c^2}\right)(v\cos\theta_0+v_0) \sin \theta_0 \: {\rm d}\theta_0\\
\label{eq:85.50}
&=&\gamma_0^2\rho_0\left(1+\frac{v^2}{3c^2}\right)\, {v_0\over c^2}=\gamma_0^2\left(\rho_0+p\right)\, {v_0\over c^2} \:.
\end{eqnarray}
where we have made use of Eq.~(\ref{eq:85.47}). 

The derived Eqs.~(\ref{eq:85.49a}) and (\ref{eq:85.50}) are the desired expressions for the energy and momentum densities in the lab frame ${\cal S}$ (Eqs.~(\ref{eq:85.4c}) and (\ref{eq:85.5c})). As was mentioned in Section 2, one would intuitively expect $\rho = \gamma_0^2\rho_0$ and $\sigma= \gamma_0^2\rho_0 v_0/c^2$. In fact this would indeed be the case if only the energy of individual molecules had gone up by a factor ${\cal D}$ without an accompanying increase in the number density by another ${\cal D}$ factor (which is much more than just a Lorentz contraction factor $\gamma$ of the volume element). It is the factor ${\cal D}^2$ in the energy density that results in a term $\propto v^2$ that survives when integrated over $\theta$ from $0$ to $\pi$, leading to the pressure term in the energy-momentum density expressions.

We can compare the effects of the magnitude of random motion in ${\cal S}_0$ and that of the bulk motion of the liquid seen in ${\cal S}$. If the random motion in ${\cal S}_0$ is 
non-relativistic ($v\ll c$), then pressure plays only a minimal part in the expressions for  energy-momentum densities (Eqs.~(\ref{eq:85.49a}) and (\ref{eq:85.50})). However, if the random motion is relativistic with $v$ comparable to $c$, then pressure too is comparable to $\rho_0$.
But even then pressure plays a substantial role in the energy density (Eq.~(\ref{eq:85.49a})) only if the bulk motion is also relativistic ($v_0 \sim c$). On the other hand, pressure plays a significant role in the momentum density (Eqs.~(\ref{eq:85.49b})) even if the bulk motion happens to be non-relativistic.

We can, of course, relax the restriction of the gas containing only one category of molecules and that too, with only a single value of speed $v$. For that we make a summation of the energy density in frame ${\cal S}_0$ over all categories of molecules in the gas, $\rho_0=\Sigma (n_0m_0\gamma)_{\rm i}\,c^2$, and $p=\Sigma n_{\rm i}m_{\rm i} \gamma_{\rm i}{v_{\rm i}^{2}}/3$, however, with a caveat that each $i$th category has an isotropic velocity distribution in frame ${\cal S}_0$. Then we still obtain expressions for $\rho$ and $\sigma$, the energy and momentum densities of the gas in the lab frame ${\cal S}$, in complete agreement with Eqs.~(\ref{eq:85.4c}) and (\ref{eq:85.5c}).
%---------------------------------------------------------
\section{A photon gas as a perfect fluid}
Let us assume that the volume, $V_0$, inside the container is filled with radiation, made up of photons of energy $h \nu_0$, having an isotropic distribution, as seen in the rest frame ${\cal S}_0$. As seen in the lab frame ${\cal S}$, the container of the fluid has a bulk motion $v_0$, say, along the x-axis. 

Now, the pressure $p$ for the radiation is related to the energy density $\rho_0$ as (see Appendix A, Eq.~(\ref{eq:85.a7}))
\begin{eqnarray}
\label{eq:85.4.1}
p=\rho_0/3\,.
\end{eqnarray}
We can then write from Eqs.~(\ref{eq:85.4c}) and (\ref{eq:85.5c}), the energy and  momentum densities for the fluid elements in the lab frame ${\cal S}$ as
\begin{eqnarray}
\label{eq:85.4a}
\rho =  \gamma_0^2 \rho_0\left(1 + {v_0^2\over 3c^2}\right)\:,
\end{eqnarray}
\begin{eqnarray}
\label{eq:85.5a}
\sigma= \frac{4}{3}\,\gamma_0^2\rho_0  {v_0\over c^2} \:.
\end{eqnarray}
Here we see that the pressure, being comparable to the energy density $\rho_0$, even in case of  a non-relativistic bulk motion of the fluid, makes a one-third contribution to the momentum density (Eq.~(\ref{eq:85.5a})), although its contribution to the energy density ($\rho$, Eq.~(\ref{eq:85.4a})), could be ignored for a non-relativistic motion, with $v_0 \ll c$. We want to derive Eqs.~(\ref{eq:85.4a}) and (\ref{eq:85.5a}) in the lab frame {\cal S}, with respect to which the fluid element has a bulk motion $v_0$.

A photon moving along $\theta_0$ with respect to the x-axis, as seen in  ${\cal S}_0$, will be moving  along $\theta$ in the lab frame ${\cal S}$, with the velocity components parallel 
%and perpendicular 
to the x-axis, given by \cite{RI06}
\begin{eqnarray}
\label{eq:85.42a}
{\displaystyle \cos \theta={\frac {\cos \theta_0+v_0/c}{1+{ {v_0\cos \theta_0}/{c}}}}\,,}\\
\label{eq:85.42b}
{\displaystyle \cos \theta-\frac{v_0}{c}={\frac {\cos \theta_0}{\gamma^2_0(1+{ {v_0\cos \theta}/{c}})}}\,.}
 \end{eqnarray}
Now a photon of energy $h\nu_0$ in  frame ${\cal S}_0$, will have in the lab frame {\cal S} an energy $h\nu=h\nu_0 \delta$, where $\delta$ is the Doppler factor given by \cite{DS98,RI06} 
\begin{eqnarray}
\label{eq:85.61}
\delta={\gamma_0(1+v_0\cos_0 \theta/c)}
\,.
\end{eqnarray}
The expression for Doppler factor $\delta$ can be obtained from {\cal D} (Eqs.~(\ref{eq:85.48})) by substitution of $V=v=c$.
 
Now as in the case of molecular gas in the Section 3, even though the motion of photons in the frame ${\cal S}_0$ has an isotropic distribution, in frame ${\cal S}$ it would not be so. This happens because a photon with a velocity component $c\cos \theta$ in frame ${\cal S}$ would take a time $\tau$ between successive collisions with two opposite container walls $AB$ and $CD$ that are normal to the x-axis, given by  
\begin{eqnarray}
\label{eq:85.62}
c\tau\cos \theta =\frac{l}{\gamma_0}+v_0\tau\,,
\end{eqnarray}
where length $l$ of the container is Lorentz contracted along the $x$ direction by a factor $\gamma_0$.
Thus the time interval between successive collisions  
\begin{eqnarray}
\label{eq:85.66}
\tau =\frac{l}{\gamma_0(c\cos \theta-v_0)}\,,
\end{eqnarray}
in frame ${\cal S}$ will be longer than $\tau_0=l/c\cos \theta_0$ in frame ${\cal S}_0$ and the corresponding photon will be spending more time between successive collision with the walls by the factor, $\cos \theta_0/{\gamma_0}(\cos \theta-v_0/c)$, before a change in $\cos \theta$ takes place because of a collision with a wall normal to the x-axis. Therefore the number density of photons moving along $\cos \theta$ will be relatively higher by a factor $\cos \theta_0/{\gamma_0}(\cos \theta-v_0/c)$ in frame ${\cal S}$, even though in frame ${\cal S}_0$ the number densities of photons with components in all directions were uniform.
\begin{eqnarray}
\label{eq:85.67}
\frac{\tau}{\tau_0}= \frac{\cos \theta_0}{\gamma_0(\cos \theta-v_0/c)}\,,
\end{eqnarray}

Using Eqs.~(\ref{eq:85.42b}) and (\ref{eq:85.61}), we can write
\begin{eqnarray}
\label{eq:85.68}
\frac{\tau}{\tau_0}=\delta \:,
\end{eqnarray}
Accordingly, the number density of photons in frame ${\cal S}$ is higher by a factor $\delta$. This is in addition to the fact that the energy of individual photons will also be higher by a  factor $\delta$ (Eqs.~(\ref{eq:85.61})).

The energy density of photon gas in ${\cal S}$ can be obtained by an integration over the solid angle, as
\begin{eqnarray}
\rho &=&\frac{n_0 h \nu_0}{4\pi}\int_{0}^{\pi} \delta^2 2\pi\sin \theta_0 \: {\rm d}\theta_0
\end{eqnarray}
\begin{eqnarray}
\nonumber
\rho &=&\frac{n_0 h \nu_0}{2}\int_{0}^{\pi} \gamma_0^2 (1+v_0\cos \theta_0/c)^2 \sin \theta_0 \: {\rm d}\theta_0\\
\label{eq:85.64}
&=&\gamma_0^2\rho_0\left(1+\frac{v_0^2}{3c^2}\right)\,,
\end{eqnarray}
where $\rho_0=n_0 h \nu_0$ is the energy density of the radiation. 

The momentum density is computed in the same way to get
\begin{eqnarray}
\nonumber
\sigma &=&\frac{n_0 h \nu_0}{4\pi c}\int_{0}^{\pi} \delta^2 \cos\theta \,2\pi\sin \theta_0 \: {\rm d}\theta_0\,,\\
\nonumber
&=&\frac{\gamma_0^2\rho_0}{2c}\int_{0}^{\pi}  \left(1+\frac{v_0\cos \theta_0}{c}\right)(\cos\theta_0+v_0/c) \sin \theta_0 \: {\rm d}\theta_0\,,\\
\label{eq:85.65}
&=&\frac{\gamma_0^2}{c}\rho_0\left(\frac{4v_0}{3c}\right)=\frac{4}{3}\,\gamma^2_0\rho_0  {v_0\over c^2} \:.
\end{eqnarray}
where we have used Eq.~(\ref{eq:85.42a}). 

The derived Eqs.~(\ref{eq:85.64}) and (\ref{eq:85.65}) are the desired expressions for the energy and momentum densities of the radiation in the lab frame ${\cal S}$ (Eqs.~(\ref{eq:85.4a}) and (\ref{eq:85.5a})). Like in the case of molecular gas in Section 3, here too not only the energy of individual photons goes up by a factor ${\delta}$, there is an accompanying increase in the number density by the same factor ${\delta}$. It is this ${\delta}^2$ factor in the energy density that results in the pressure term in the energy-momentum density expressions.

In the case of radiation, pressure is comparable to the energy density in ${\cal S}_0$ ($p=\rho_0/3$). But even here pressure plays only a minimal part in the expressions for the energy density (Eq.~(\ref{eq:85.4a})) if the bulk motion is non-relativistic ($v_0 \ll c$). However, pressure does play a significant role in the momentum density (Eqs.~(\ref{eq:85.5a})) even if the bulk motion happens to be non-relativistic.
The expressions (Eqs.~(\ref{eq:85.49a}) and (\ref{eq:85.50})) for the energy and momentum densities in the case of an ultra-relativistic gas ($v\approx c$) reduce to those of the radiation (Eqs.~(\ref{eq:85.64}) and (\ref{eq:85.65})).

In the case the photon gas has a frequency distribution, e.g., in a Planckian distribution, then $\rho_{0}=\Sigma (n_0 \nu_0)_{\rm i}\, h$, again assuming that photons of each $i$th frequency have an isotropic velocity distribution. Then we still obtain expressions for $\rho$ and $\sigma$, the energy and momentum densities of the radiation in the lab frame ${\cal S}$, in complete agreement with Eqs.~(\ref{eq:85.4a}) and (\ref{eq:85.5a}).
%---------------------------------------------
\section{Conclusions}
We derived, from first principles, expressions for the energy and momentum densities of an ideal gas,  which is a simple example of a perfect fluid, in an inertial frame in which the fluid moves, with a constant bulk velocity. Our Starting point was the rest frame, where the random motion of the gas constituents, giving rise to the pressure, is isotropic and the fluid accordingly has no bulk motion, and where stress-energy tensor is of a diagonal form, described by only the energy density and pressure. From a Lorentz transformation of energy and momentum of individual molecules, moving along different directions in the rest frame of the fluid, and the transformation of their number densities, we computed the energy-momentum density of the fluid in the frame where it has a bulk motion. We thereby showed how a bulk motion of the fluid, in the presence of random motion in the rest frame within the fluid, gives rise to a couple of pressure-dependent terms in the energy and momentum densities of a perfect fluid system, which is an ideal gas in our case. We also derived relations for the energy-momentum density of radiation in terms of the energy density and pressure of the photon gas in the frame where the gas has an isotropic distribution.
%---------------------------------------------------------------------------
\begin{appendix}
%--------------------------------------------------------------------
\section{Pressure in an ideal gas model of a perfect fluid}
We assume that the perfect fluid comprises an ideal gas containing, for simplicity, a homogeneous number density $n_0$ of identical, non-interacting, structureless molecules, each of rest mass $m_0$, and moving with the same speed $v$, though in random directions.

The fluid may be confined within the walls of a container, supposed to be stationary in an inertial frame, say ${\cal S}_0$. The distribution of velocities of molecules is thus assumed to be isotropic in the rest frame ${\cal S}_0$. We further assume that the molecules of the gas are ``point masses'' that may undergo only elastic collisions with each other and the walls of the container in which both linear momentum and kinetic energy are conserved. The elastic collisions may give rise to a finite pressure, but no exchange of energy, thus no heat conduction. We also assume the molecules to be otherwise non-interacting, so no ``sideways'' interactions with other molecules in the neighbourhood thus no sheer stress, no drag or no viscosity. 

A molecule, initially moving along an angle $\theta_0$ with respect to the x-axis, on encountering a container wall, say, $CD$ (Fig. 1), normal to the x-axis, impinges upon it with a velocity component $v_{\rm x}=v \cos \theta_0$, rebounds with a velocity component $-v_{\rm x}$, and thereby impart a momentum to the wall along $x$ direction
\begin{eqnarray}
\label{eq:85.a1}
{\displaystyle \Delta P=2\gamma m_0 v \cos \theta_0}\,.
\end{eqnarray}
where $v$ may be relativistic with a Lorentz factor $\gamma=(1-v^2/c^2)^{-1/2}$.

The number of molecules that hit a unit area of the wall $CD$, while moving along $\theta_0$, per unit solid angle per second, is $n_0 v \cos \theta_0/4\pi$. The pressure is force per unit area, exerted by the molecules of the gas on the walls of the stationary container. Assuming an isotropic distribution of velocity vectors, the pressure of the gas on the wall $CD$ is
\begin{eqnarray}
\label{eq:85.a2}
    {\displaystyle p={\frac {2n_0m_0\gamma v^2}{4\pi}\int_{0}^{\pi/2} \cos^2\theta_0 \,2\pi \sin \theta_0 \: {\rm d}\theta_0\,}.}
\end{eqnarray}
It should be noted that molecules moving only within $0\le\theta < \pi/2$ will hit the wall $CD$. The pressure can thus be written as
\begin{eqnarray}
\label{eq:85.a4}
    {\displaystyle p={\frac {n_0m_0\gamma{ {v^{2}}}}{3}}
    ={\frac {\rho_0{ {v^{2}}}}{3c^2}}
\,.}
\end{eqnarray}
Here $\rho_0=n_0m_0 c^2\gamma$ is the relativistic energy density of the gas, that includes the rest mass energy as well as the kinetic energy of molecules per unit volume. 

We can relax the restriction of the gas containing only one category of molecules and that too, with only a single value of speed $v$. In such cases, we sum the energy density over all categories of molecules in the gas, having even some spread in the energy distribution, $\rho_0=\Sigma n_{\rm i}m_{\rm i} \gamma_{\rm i}\,c^2$, and $p=\Sigma n_{\rm i}m_{\rm i} \gamma_{\rm i}{v_{\rm i}^{2}}/3$, where each $i$th category has an isotropic velocity distribution. 

The expression for pressure in ultra-relativistic gas ($v\approx c$) simply becomes 
\begin{eqnarray}
\label{eq:85.a8}
    {\displaystyle p={\frac {\rho_0}{3}}\,.}
\end{eqnarray}

If the random motion of the molecules is non-relativistic, with $v\ll c$ and $\gamma \approx 1$, then the pressure can be written as
\begin{eqnarray}
\label{eq:85.a5}
    {\displaystyle p={\frac {n_0m_0{ {v^{2}}}}{3}}    =\frac {2\rho_{\rm k}}{3}
\,.}
\end{eqnarray}
where ${\rho_{\rm k}}=n_0m_0 {v^{2}}/2$ is the kinetic energy density of the molecules. In the case the gas contains more than one category of molecules and with their speed distributions having a certain spread, e.g., a Maxwellian distribution of the air at a room temperature, then $\rho_{\rm k}=\Sigma n_{\rm i} m_{\rm i} v^2_{\rm i}\,/2$, again assuming each $i$th category has an isotropic velocity distribution. 

On the other hand, if the container comprises radiation, then we can replace the gas molecules by photons with a number density $n_0$, each of energy, say $h \nu_0$, moving in random directions. The change in the momentum of a photon, initially moving along $\theta_0$ with respect to the x-axis, after collision with the container wall, is
\begin{eqnarray}
\label{eq:85.a6}
{\displaystyle \Delta P=\frac{2h \nu_0 \cos \theta_0}{c}}\,,
\end{eqnarray} 
while the number of photons moving along $\theta_0$, per unit solid angle, reaching the wall per second is $n_0 c \cos \theta_0/4 \pi$, implying a pressure due to an isotropic distribution of photon velocities
\begin{eqnarray}
\label{eq:85.a7}
    {\displaystyle p={ \frac{2n_0h \nu_0}{4\pi}\int_{0}^{\pi/2} \cos^2\theta_0 \, 2\pi\sin \theta_0 \: {\rm d}\theta_0\,}={\frac {\rho_0}{3}}}\,,
\end{eqnarray}
where $\rho_0=n_0 h \nu_0$ is the energy density of the radiation.
In the case the photons are with a spread in the frequency distribution, e.g., in a Planck distribution, then $\rho_{0}=\Sigma n_{\rm i} h \nu_{\rm i}$, again assuming photons of each $i$th frequency having an isotropic velocity distribution. 

Of course, the expression for the radiation pressure is the same as for the ultra-relativistic gas (Eq~(\ref{eq:85.a8})).
%---------------------------------------------------------------
\section{Pressure of an ideal gas model of perfect fluid with a bulk motion}
Conventionally, pressure is defined as force per unit area on a fluid cross-section, in the rest frame ${\cal S}_0$, where the fluid element has no bulk motion. We assume that the fluid moves with a bulk velocity $v_0$ with respect to the frame ${\cal S}$, implying the rest frame ${\cal S}_0$ is moving with  velocity $v_0$ with respect to the frame ${\cal S}$. Then pressure,  defined as force per unit area, can be calculated in the frame ${\cal S}$ too, where the fluid has a bulk motion, however, for that we have to choose a cross-section, say, $CD$, that is stationary in ${\cal S}_0$ (Fig.~1), and has a motion with respect to the frame ${\cal S}$ (Fig.~2).  

Now for the ideal, relativistic, gas model of Appendix A, we want to calculate the force exerted per unit area on the wall $CD$ that is moving with velocity $v_0$ along x-axis, as seen in frame ${\cal S}$ (Fig.~2). A molecule moving with velocity, $v$, and a velocity component $v_{\parallel}=v\cos\theta_0$, along the x$_0$-axis, in ${\cal S}_0$, after an elastic  collision with the wall $CD$, bounces off with a velocity component $-v_{\parallel}$ along the x$_0$-axis. This molecule, as seen in the lab frame ${\cal S}$, will be moving, before and after the collision, respectively with velocities $V^+_{\parallel}$, $V^-_{\parallel}$, as
\begin{eqnarray}
\nonumber
{\displaystyle V^+_{\parallel}={\frac {v_{\parallel}+v_0}{1+{ {v_{\parallel}v_0}/{c^{2}}}}},}\\
\label{eq:85.a12}
{\displaystyle V^-_{\parallel}={\frac {-v_{\parallel}+v_0}{1-{ {v_{\parallel}v_0}/{c^{2}}}}}\,,}
 \end{eqnarray}
and the corresponding Lorentz factors
\begin{eqnarray}
\nonumber
\Gamma^+=\gamma_0\gamma\left(1+\frac{v_0v_{\parallel}}{c^2}\right)\,,\\
\label{eq:85.a13}
\Gamma^-=\gamma_0\gamma\left(1-\frac{v_0v_{\parallel}}{c^2}\right)\,.
\end{eqnarray}
Therefore, the impact on the wall $CD$, due to a collision of an individual molecule, will be
\begin{eqnarray}
\nonumber
\Delta P&=&m_0(\Gamma^+V^+_{\parallel}-\Gamma^-V^-_{\parallel}),\\
\label{eq:85.a14}
&=&2m_0\gamma_0\gamma v_{\parallel}\,.
\end{eqnarray}
The number of molecules with velocity $V^+_{\parallel}=V\cos\theta$ hitting the wall $CD$ (moving with velocity $v_0$ in ${\cal S}$), per unit of time, 
%in the solid angle ${\rm d}\Omega=\sin \theta{\rm d}\theta{\rm d}\phi$, 
in the frame ${\cal S}$ will be proportional to 
\begin{eqnarray}
%\nonumber
\Delta {\cal N} &\propto&  {\cal D} (V\cos\theta-v_0)\,,
\end{eqnarray}
the factor ${\cal D}$ here is due to the increase in number density in the frame ${\cal S}$.

%By integrating over the solid angle,
% with a change of variable from $\theta$ to $\theta_0$ (Eqs.~(\ref{eq:85.47}) and (\ref{eq:85.47a})),  
The net force per unit area or pressure, measured in frame ${\cal S}$, in a direction parallel to the bulk motion of the fluid, then is obtained from an integration over the solid angle.
\begin{eqnarray}
\nonumber
p &=&\frac{2m_0\gamma n_0}{4\pi}\int_{0}^{\pi/2}\gamma^2_0 \left(1+\frac{v_0v \cos \theta_0}{c^2}\right) v_{\parallel} (V\cos\theta-v_0)\, 2\pi \sin \theta_0 \: {\rm d}\theta_0\\
\nonumber
&=&{\frac {\rho_0{ {v^{2}}}}{c^2}}\int_{0}^{\pi/2}  \cos^2\theta_0 \sin \theta_0 \: {\rm d}\theta_0\\
\label{eq:85.a15}
&=& {\frac {\rho_0{ {v^{2}}}}{3c^2}}\,,
\end{eqnarray}
where we have used Eq.~(\ref{eq:85.47.1}), and $\rho_0=n_0m_0 c^2\gamma$ is the relativistic energy density of the gas. The value for pressure $p$ in frame ${\cal S}$ calculated this way is consistent with that in the rest frame ${\cal S}_0$ (Eq~(\ref{eq:85.a4})).

For calculating the pressure in a direction normal to the bulk flow, we consider a molecule moving with velocity, $v$, and velocity components $v_{\parallel},v_{\perp}$ in  ${\cal S}_0$. 
%after a collision 
%towards the wall $BD$.
% bounces off with a velocity component $v_{\parallel},-v_{\rm y}$. 
This molecule will be moving in the lab frame ${\cal S}$ with a velocity components $V_{\perp}$ \cite{KKR73} 
\begin{eqnarray}
%\nonumber
%{\displaystyle V^+_{\parallel}={\frac {v_{\parallel}+v_0}{1+{ {v_{\parallel}v_0}/{c^{2}}}}},}\\
\label{eq:85.a16}
 {\displaystyle V_{\perp}={\frac {v_{\perp}}{\gamma_0({1+{ {v_{\parallel}v_0}/{c^{2}}}})}}=\frac{v_{\perp}}{\cal D}\,.}
 \end{eqnarray}
After the collision, $v_{\perp}$  will be reversed in sign, and so will be $V_{\perp}$. Therefore the momentum imparted to a wall normal to the $x$ direction by this molecule will be
\begin{eqnarray}
\label{eq:85.a17}
\Delta P=2m_0V_{\perp}\Gamma^+=2m_0v_{\perp}\gamma\,.
 \end{eqnarray}
The number of molecules with velocity $V_{\perp}$, hitting a wall normal to the $x$ direction, per unit of time, 
%in the solid angle ${\rm d}\Omega=\sin \theta{\rm d}\theta{\rm d}\phi$, 
in the frame ${\cal S}$ will be proportional to
\begin{eqnarray}
\label{eq:85.a18}
\Delta {\cal N} &\propto&  {\cal D} V_{\perp}\,,
\end{eqnarray}
where ${\cal D}$ is due to the increase in number density in the frame ${\cal S}$. 
It should be noted that molecules in only one fourth of azimuth angle range, i.e., $\pi/2$, will be hitting any one particular wall, say, $BD$ (Fig.~2), normal to the direction of the bulk motion.

Integrating over the solid angle, the net force per unit area or pressure measured in frame ${\cal S}$, in a direction normal to the bulk motion of the fluid, then is
\begin{eqnarray}
\nonumber
p &=&\frac{2m_0\gamma n_0}{4\pi}\int_{0}^{\pi} {\cal D} V_{\perp} v_{\perp} \, \frac{\pi}{2} \sin \theta_0 \: {\rm d}\theta_0\,,\\
\nonumber
&=&\frac{m_0 n_0\gamma}{4}\int_{0}^{\pi}  v^2_{\perp} \sin \theta_0 \: {\rm d}\theta_0\,,\\
\nonumber
&=&\frac{m_0 n_0\gamma}{4}{v^{2}}\int_{0}^{\pi}  \sin^2\theta_0 \sin \theta_0 \: {\rm d}\theta_0\\
\label{eq:85.a19}
&=& {\frac {\rho_0{ {v^{2}}}}{3c^2}} 
%    {\displaystyle p={n_0m_0\gamma{  \overline{v_{\rm y}^{2}}}}
\end{eqnarray}
which is the same as in the frame ${\cal S}_0$ (Eq~(\ref{eq:85.a4})). 
Thus the pressure is an invariant quantity.

In the case of radiation, we can put $v=V=c$ in the above formulation, which will also make ${\cal D}=\delta$, and we get 
\begin{eqnarray}
\label{eq:85.a20}
    {\displaystyle p={\frac {\rho_0}{3}}
\,.}
\end{eqnarray}
in all directions. 

It is important to note that in Eq.~(\ref{eq:85.a20}) (or even in (\ref{eq:85.a19})), pressure $p$ is not derived in terms of $\rho$, the energy density in the frame ${\cal S}$ where the fluid element has a bulk motion, i.e., $p\ne{\rho}/{3}$ in the case of radiation (or $p\ne{\rho}V^2/{3c^2}$ for gas comprising molecules), and that the pressure instead is defined properly in terms of the energy density $\rho_0$ in frame ${\cal S}_0$, where the random motion of the gas constituents is of an isotropic distribution.
%---------------------------------------------------------------
\end{appendix}
%--------------------------------------------------------------------
%\newpage
{}
\end{document}